\documentclass[
twocolumn,
preprintnumbers,
nofootinbib,
prd,aps
]{revtex4}

\usepackage{latexsym}
\usepackage{amssymb}
\usepackage{epsf}
\usepackage{amsmath}
\usepackage{slashed}
\usepackage{docmute}
 \usepackage{graphicx}
 \usepackage{hyperref}
\usepackage{bm}
\usepackage{cases}
\usepackage{amssymb}
\usepackage{subfigure}
\usepackage{subfiles}
\usepackage{amsfonts}

\begin{document}


\newcommand{\rem}[1]{$\spadesuit${\bf #1}$\spadesuit$}

\renewcommand{\topfraction}{0.8}

\preprint{OU-HET 1057}

\title{Higgs boson coupling as a probe of the sphaleron property}

\author{Shinya Kanemura and Masanori Tanaka}

\affiliation{Department of Physics, Osaka University,Toyonaka, Osaka 560-0043, Japan}


\begin{abstract}

Sphaleron is a non-perturbative solution of electroweak gauge theories, which is crucially important for the scenario of electroweak baryogenesis. The sphaleron energy depends on details of the mechanism for the electroweak symmetry breaking.
We find that in many of new physics models the deviation of the sphaleron energy from the prediction in the standard model is proportional to that of the triple Higgs boson coupling with opposite signs. This interesting relation would be useful to determine the sphaleron energy  by measuring this couplings at future collider experiments.

\end{abstract}

\maketitle

\renewcommand{\thefootnote}{\#\arabic{footnote}}

\section{Introduction}

Although the standard model (SM) for particle physics is successful, the origin of the electroweak symmetry breaking (EWSB) remains unknown.
In spite of the discovery of the Higgs boson $h(125)$ in 2012, the structure of the Higgs sector is still mysterious.
In addition, there are several important phenomena which cannot be explained in the SM,
such as neutrino oscillation, dark matter and baryon asymmetry of the Universe (BAU).
It is obvious that a new model beyond the SM is necessary.

Baryogenesis is a concept of new physics beyond the SM to explain BAU from the baryon symmetric initial state.
In promising scenarios for baryogenesis, either leptogenesis\cite{Fukugita:1986hr} or electroweak (EW) baryogenesis\cite{Kuzmin:1985mm},
the sphaleron transition plays a crucial role,
which can change the baryon number at finite temperatures of the early Universe.
Sphaleron is a non-perturbative solution of the EW gauge theory.
In the $SU(2)$ gauge theory, the sphaleron solution was first studied by Manton\cite{Manton:1983nd},
and its extension to the $SU(2)\times U(1)$ symmetry, i.e., the EW gauge symmetry of the SM,
was investigated by Klinkhamer and Manton\cite{Klinkhamer:1984di}.
Their works have been followed by Refs.~\cite{Akiba:1988ay,Kleihaus:1991ks}.
The energy of the sphaleron is evaluated as about 9.1TeV in the SM with an isospin-doublet scalar field (the Higgs field).

On the other hand, many viable new physics models require non-standard Higgs sectors,
such as multi-scalar extensions like supersymmetric models and composite structures for the nature of the Higgs field.
These models also have the sphaleron solution, as they contain the structure of the EW gauge theory
\cite{Kastening:1991nw,Funakubo:2009eg,Spannowsky:2016ile}.
Their sphaleron property can be different from the one in the SM due to the different Higgs sectors.
In particular, the sphaleron energy takes specific values for each model.
As Spannowsky and Tamarit investigated\cite{Spannowsky:2016ile}, lower sphaleron energies
than that in the SM are predicted in models with a deformed Higgs potential, while
higher ones are predicted in minimal composite Higgs models (MCHMs).
The sphaleron energy was also studied in models with dimension-six operators by Gan et al.~\cite{Gan:2017mcv}.

The sphaleron energy depends on new physics models, so that the model may be able to be discriminated
by the sphaleron energy. Then, how can we determine the sphaleron energy by experiments?
Several studies have been performed to detect the sphaleron at future collider experiments or astronomical
observations\cite{Ellis:2016ast}.
The possibility of direct detection of the sphaleron at future high-energy colliders has been proposed in Ref.~\cite{Tye:2015tva}, while
the solidness of the calculation is still under discussion\cite{Funakubo:2016xgd}.

In this Letter, we shall discuss a new method to determine the sphaleron energy in new physics models at future experiments.
The sphaleron energy is affected by the shape and the structure of the Higgs potential.
This would suggest that by measuring the Higgs potential at future experiments we can determine the sphaleron energy,
by which we can narrow down the direction of new physics beyond the SM.
The deformed Higgs potential can affect the triple Higgs boson coupling of $h(125)$,
which is expected to be measured at the High-Luminosity LHC\cite{ATLAS:2013hta} and more precisely at future high-energy lepton\cite{Fujii:2019zll,CLIC:2016zwp} or hadron colliders\cite{Abada:2019ono}.

The deviation in the sphaleron energy from the SM prediction is found to be proportional
to that in the triple Higgs boson coupling with opposite sign in a good approximation in various extended Higgs models; i.e.,
the model with a dimension-six operator in the Higgs potential,
the renormalizable models with additional scalar fields in the Higgs sector,
those with the classical scale invariance (CSI),
and MCHMs with several matter representations.
Using the relation with the well known value of the sphaleron energy of the SM, we can empirically determine the sphaleron energy
in new physics models by measuring the deviation in the triple Higgs boson coupling.
There have been only few studies to test the sphaleron property by experiments.
Therefore, our new results are important to obtain the information on the sphaleron property in various new physics models.

\section{Sphaleron in the standard model}

We begin our discussion with the sphaleron energy in the SM.
As shown in Ref.~\cite{Kleihaus:1991ks}, the effect of the $U(1)$ hypercharge
does not change much the field configuration of the sphaleron in the $SU(2)$ gauge theory.
For the deviation in the sphaleron energy from the SM prediction due to the
non-minimal Higgs sector, we simply discuss the $SU(2)$ gauge theory\cite{Manton:1983nd}.
We also do not consider the fermions in this Letter, as their effect on the sphaleron energy is usually negligible\cite{Nolte:1993jt}.

The energy functional is defined by
\begin{eqnarray}
&&E[W_i^a, \Phi] =\nonumber \\
&& \int d^3 x \left(
\frac{1}{4g^2} W^a_{ij} W^{a ij} + ({\cal D}_i \Phi)^\dagger {\cal D}^i \Phi + V(\Phi)
\right),
\label{eq:sph_energy}
\end{eqnarray}
where $W^a_{ij}$ is the field strength of the $SU(2)$ gauge field $W^a_i$ with $a = 1$-$3$ being the $SU(2)$ index and $i,j=1$-$3$ being the index
of the space component of the field, $g$ is the gauge coupling constant, and $\Phi$ is the isospin-doublet Higgs field.
We have imposed that the field configuration is static, and the time component of the gauge field is zero.
In the SM, the covariant derivative and the Higgs potential are given by
\begin{eqnarray}
 {\cal D}_i = \partial_i - i g\frac{\sigma^a}{2}  W^a_i, \;V^{\rm SM}(\Phi) = \frac{m_h^2}{2 v^2} \left(\Phi^\dagger \Phi - \frac{v^2}{2}\right)^2,
\end{eqnarray}
where $\sigma^a$ represent the Pauli matrices, $v$  ($\simeq 246$GeV) is the vacuum expectation value (VEV)  of $\Phi$, $m_h$ ($\simeq 125$GeV) is the physical mass of the Higgs boson $h$.

\begin{figure}[t]
  \begin{center}
    \includegraphics[width=6cm]{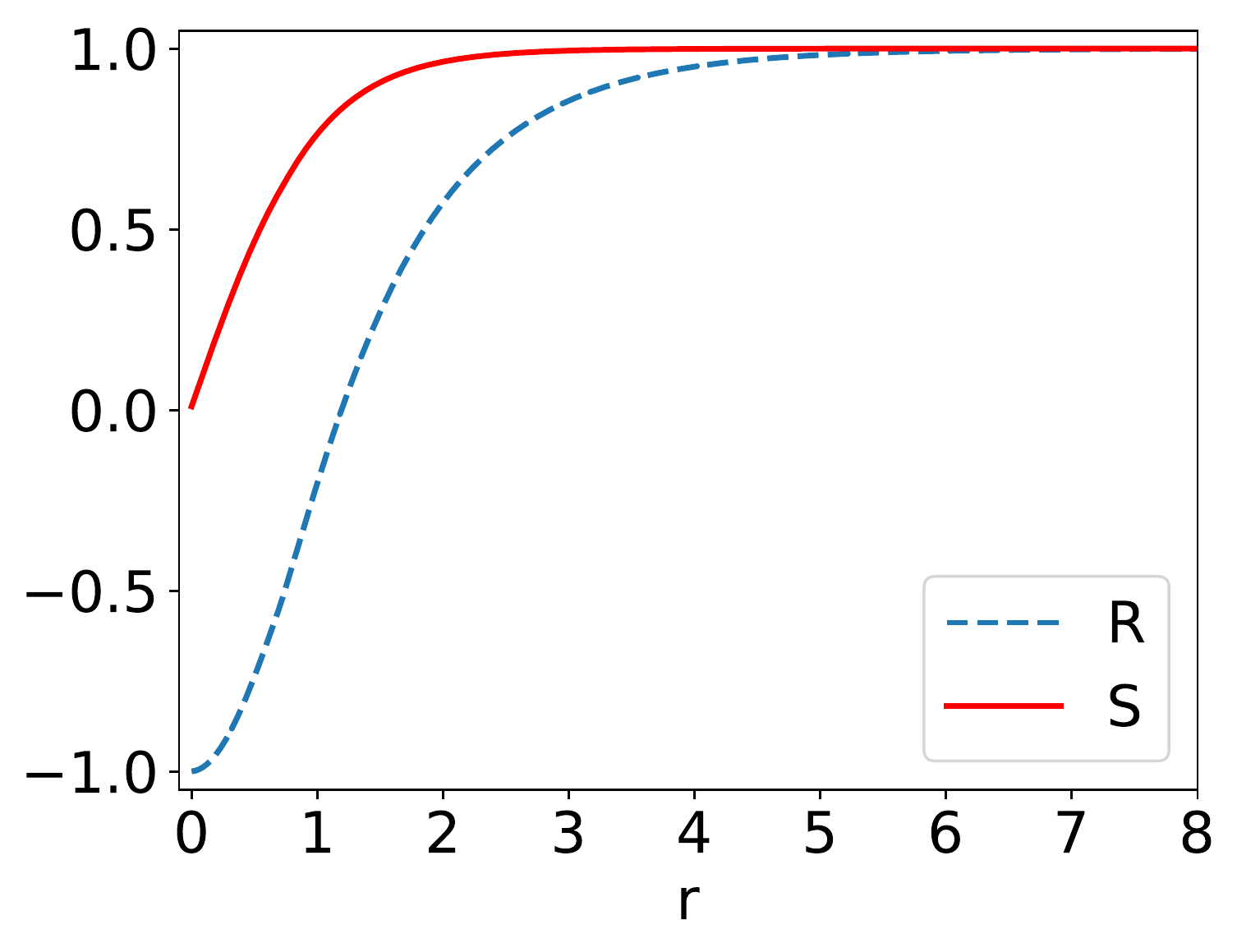}
    \caption{The values of the functions $R(r)$ and $S(r)$ in the SM.}
    \label{fig:SR}
  \end{center}
\end{figure}

Using the hedgehog ansatz\cite{Akiba:1988ay},
the sphaleron solution can be expressed in terms of functions
$R(r)$, $S(r)$, $\theta(r)$ and $\phi(r)$ by
\begin{eqnarray}
W^a_i(r)&=& m_W^{}\left[ \epsilon^{aij} n_j \frac{1-R(r)\cos\theta(r)}{r}\right.\nonumber \\&&\left.+(\delta_{ai}-n_a n_i) \frac{R(r)\sin\theta(r)}{r}\right],\\
\Phi(r) &=& \frac{v}{\sqrt{2}} S(r) e^{in_a \sigma^a \phi(r)} \left(\begin{array}{c} 0\\ 1\\ \end{array} \right),
\end{eqnarray}
where $n_a = x^a/r$.
We take $\theta(r)$ and $\phi(r)$ so as to satisfy $\theta'=\phi'=0$ and $\phi=\theta/2+ \omega\pi/2$ ($\omega \in Z$)\cite{Spannowsky:2016ile}.
Consequently, the field equations become
\begin{eqnarray}
&&\hspace*{-6mm}r^2 R'' - R^3 + R(1-r^2 S^2) \pm r^2 S^2=0, \label{eq:r}\\
&&\hspace*{-6mm}2r^2 S'' + 4r S' - S \left\{ (R\mp 1)^2 + \kappa^2 r^2(S^2-1) \right\}=0,\label{eq:s}
\end{eqnarray}
where $\kappa = m_h^{}/m_W^{}$.
In order to obtain physically allowed solutions, we take $\omega$ to be even,
which corresponds to the upper sign in Eqs.~(\ref{eq:r}) and (\ref{eq:s}).
We numerically obtain $R(r)$ and $S(r)$ as shown in Fig.~\ref{fig:SR}.
Finally the sphaleron energy is found to be $E_{\sf sph}^{\rm SM} = 9.08$TeV.

\section{Sphaleron in new physics models}

We now turn to the sphaleron energy in new physics models.
In order to satisfy the current LHC data, these models should have the similar property to the SM
at the EW scale.  Therefore, in a good approximation, the EW symmetry is broken
by the VEV of only one SM-like Higgs field.
In this case, the sphaleron energy can be evaluated by using Eq.~(\ref{eq:sph_energy}).
The sphaleron energy can be different from the SM value due to
the deformed shape of the Higgs potential, that of the covariant derivative term
and the different field configurations.
However, the last effect can be neglected when the vacuum expectation value is carried by one of the scalars.
The field configuration of the sphaleron in the SM can be used as long as the deviation is not too large.
We have confirmed the validity of the above statement in the models treated in this letter.
In the following, we use the field configuration of the SM (Fig. \ref{fig:SR}) as a good approximation.
The sphaleron energy in a new physics model can be written as
\begin{eqnarray}
E_{\sf sph}^{\rm new}  = E_{\sf sph}^{\rm SM}  + \Delta E_{\sf sph}^{\rm new}.
\end{eqnarray}
The deviation $\Delta E_{\sf sph}^{\rm new}$
depends on the model\cite{Spannowsky:2016ile}.

In the following, we show that $\Delta E_{\sf sph}^{\rm new}$
can be directly related to the deviation in the prediction on
the triple Higgs boson coupling in various new physics models;
\begin{eqnarray}
\Delta E_{\sf sph}^{\rm new} = - A^{\rm new} \frac{\Delta \lambda_{hhh}^{\rm new}}{\lambda_{hhh}^{\rm SM}},
\label{eq:relation}
\end{eqnarray}
where $\Delta \lambda_{hhh}^{\rm new}= \lambda_{hhh}^{\rm new}-\lambda_{hhh}^{\rm SM}$
with $\lambda_{hhh}^{\rm new}$ being the triple coupling of $h(125)$
in the new physics model ($\lambda_{hhh}^{\rm SM}=3m_h^2/v$).
The coefficient $A^{\rm new}$ is a positive number, which depends on the model but is predictable in each model.

First, we consider the simplest example, where we just add a dimension-six operator
to the Higgs potential~\cite{Gan:2017mcv,Grojean:2004xa},
\begin{eqnarray}
 V^{\rm dim\,6}(\Phi) = V^{\rm SM}(\Phi) + \frac{1}{\Lambda^2} \left(\Phi^\dagger\Phi-\frac{v^2}{2}\right)^3,
\end{eqnarray}
where $\Lambda$ is a dimensionful parameter. The difference in the sphaleron energy from the
SM value comes from the additional term,
which is given by
\begin{eqnarray}
\Delta E_{\sf sph}^{\rm dim\,6} =
- \frac{\pi v^2 \kappa^2}{4 m_W^{}} \frac{2 v^4}{m_h^2 \Lambda^2} \int_0^\infty dr r^2(1-S(r))^3, \nonumber
\end{eqnarray}
where $S(r)$ are the SM field configuration given in Fig.~\ref{fig:SR}.
We have confirmed that the difference of the field configuration from the SM one is negligible for $\Lambda > 600$GeV.
On the other hand, the triple Higgs boson coupling can be calculated
in terms of $v$, $m_h$ and $\Lambda$ as
\begin{eqnarray}
\frac{\Delta\lambda_{hhh}^{\rm dim\,6}}{\lambda_{hhh}^{\rm SM}} = \frac{2v^4}{m_h^2 \Lambda^2}.
\end{eqnarray}
Therefore, we obtain the relation in Eq.~(\ref{eq:relation}) with
\begin{eqnarray}
A^{\rm dim\,6} = \frac{\pi v^2 \kappa^2}{4 m_W^{}}  \int_0^\infty dr r^2(1-S(r))^3 = 0.149{\rm TeV}.
\label{eq:coef_dim6}
\end{eqnarray}
We note that $A^{\rm dim\,6}$ does not depend on $\Lambda$.
For $\Lambda=600$GeV ($700$GeV), $\Delta \lambda^{\rm dim\,6}_{hhh}/\lambda_{hhh}^{\rm SM}=+130.5$\% ($+95.9\%$)
which corresponds to $\Delta E_{\sf sph}^{\rm dim\,6}/E_{\sf sph}^{\rm SM}=-2.13$\% ($-1.54\%$).

\begin{figure}[t]
  \begin{center}
    \includegraphics[width=8cm]{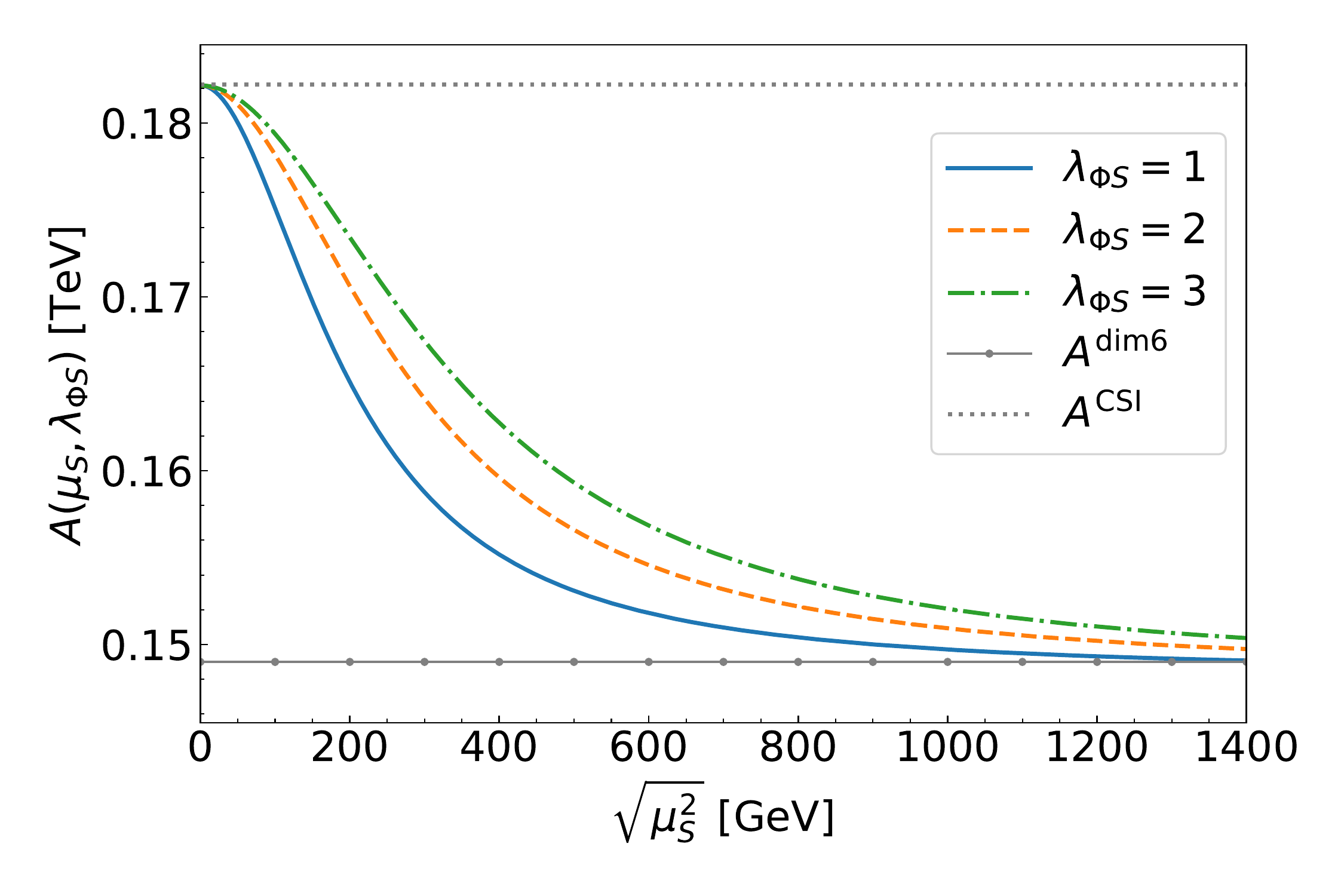}
    \caption{The coefficient $A(\mu_S^2,\lambda_{\Phi S})$ in the $N$-scalar model, together with
             $A^{\rm dim\; 6}$ and $A^{\rm CSI}$ ($= A^{\rm ND}$).}
    \label{fig:ONsinglet_coefficient}
  \end{center}
\end{figure}

Second, we discuss a simple example of the renormalizable model, in which $N$
additional isospin-singlet real scalar fields $S_i$ ($i=1$-$N$) are added to the SM.
For simplicity, we impose a global $O(N)$ symmetry, so that they have a common mass.
The Higgs potential is given by
\begin{eqnarray}
V^{N\,\rm{scalar}} = V^{\rm SM} + \Delta V^{N\,\rm{scalar}},
\end{eqnarray}
where
\begin{eqnarray}
\Delta V^{N\,\rm{scalar}}(\Phi, S_i)
= \frac{\mu_S^2}{2}  |\vec{S}|^2 + \frac{\lambda_S^{}}{4!} |\vec{S}|^4+\lambda_{\Phi S} |\vec{S}|^2 \Phi^\dagger\Phi, \nonumber\\
\end{eqnarray}
with $\vec{S}=(S_1, \cdots ,S_N)$ being a vector under the $O(N)$ symmetry.
For $\mu_S^{2} > 0$, we have $\langle S_i \rangle=0$. The
common physical mass of the scalar components $S_i$ is given by
\begin{eqnarray}
M_S^2 = \mu_S^2 + \lambda_{\Phi S}^{} v^2.
\end{eqnarray}
Including the one-loop effects, we obtain $\Delta V_{\rm eff}^{N\,\rm{scalar}}$ as
\begin{eqnarray}
&& \Delta V_{\rm eff}^{N\,\rm{scalar}}= -\frac{N \lambda_{\Phi S}^2 v^4}{64 \pi^2}\left[
  {\cal M}^2( \gamma -2 - 2 \gamma F[{\cal M}^2])  \right.\nonumber\\
&&  \left.+  {\cal M}^4 \left(\frac{3}{2} -  F[{\cal M}^2] \right)
 - \gamma^2  F[{\cal M}^2] + \left( \frac{1}{2}- \gamma \right)
  \right], \nonumber \\
\end{eqnarray}
where ${\cal M}=\sqrt{2\Phi^\dagger\Phi}/v$ and $\gamma=\mu_S^2/(\lambda_{\Phi S}^{} v^2)$, and
\begin{eqnarray}
  F[{\cal M}] = \log \frac{\mu_S^2 + \lambda_{\Phi S} v^2 {\cal M}}{\mu_S^2 + \lambda_{\Phi S}v^2}.
\end{eqnarray}
Using the hedgehog ansatz, we obtain
\begin{eqnarray}
&& \hspace{-2mm} \Delta E_{\sf sph}^{N\,\rm{scalar}} =
  -\frac{N \lambda_{\Phi S}^2 v^4}{16 \pi m_W^3} \int_0^\infty dr r^2
  \left[ S^2 (\gamma -2 - 2 \gamma F[S^2]) \right. \nonumber\\
&& \hspace{-2mm} \left.+ S^4 \left(\frac{3}{2}-F[S^2] \right) -\gamma^2
  F[S^2] + \left( \frac{1}{2}-\gamma \right) \right],
\end{eqnarray}
where the field configuration for the SM sphaleron $S=S(r)$ shown in Fig.~\ref{fig:SR} is taken as a good approximation.
For example, in the case of ($\mu_S,N$)=(0,4), we can use the field configuration of the sphaleron in the SM approximately up to $\lambda_{\Phi S}\simeq 2.4$ ($\Delta \lambda_{hhh}^{N {\rm scalar}}/\lambda_{hhh}^{\rm SM}<73\%,M_S<381$GeV).
As the value of $\mu_S$ increases, the allowed region of $\lambda_{\Phi S}$ also expands where we can use the approximation.
On the other hand, the deviation in the renormalized triple Higgs boson coupling is also
evaluated as\cite{Kakizaki:2015wua}
\begin{eqnarray}
 \frac{\Delta \lambda_{hhh}^{N\,\rm{scalar}}}{\lambda_{hhh}^{\rm SM}} =\frac{N\lambda^3_{\Phi S}v^4}{12 \pi^2 m_h^2 M_S^2}.
\end{eqnarray}
Therefore, we obtain
\begin{eqnarray}
\Delta E_{\sf sph}^{N\,\rm{scalar}} = - A (\mu_S^2, \lambda_{\Phi S}^{}) \frac{\Delta \lambda_{hhh}^{N\,\rm{scalar}}}{\lambda_{hhh}^{\rm SM}},
\end{eqnarray}
where $A(\mu_S^2, \lambda_{\Phi S}^{})$ is independent of $N$. See Fig.~\ref{fig:ONsinglet_coefficient}.

For the case of $\mu_S^2 \gg v^2$, the physical mass $M_S^{}$ becomes
large and the effect of the fields $S_i$ decouples from the low energy observables.
The mass $M_S^{}$ $(\simeq \sqrt{\mu_S^{2}})$ can be regarded as the cutoff scale $\Lambda$ of the SM.
On the other hand, for $\mu_S^2 \sim v^2$, the loop effect of the fields $S_i$ does not
decouple and the triple Higgs boson coupling is largely enhanced by
a significant non-decoupling effect of the fields $S_i$\cite{Kanemura:2002vm,Kakizaki:2015wua}.
In this case, the EW phase transition is strongly first order.
The limiting values are given by
\begin{eqnarray}
A(\mu_S^2, \lambda_{\Phi S}) \to
\left\{ \begin{array}{c}  A^{\rm dim\,6}  \;\;(\mu_S^2 \gg v^2),\\
                          A^{\rm ND} \;\; (\mu_S^2 \to 0), \\
\end{array} \right.
\end{eqnarray}
where $A^{\rm dim\,6}$ is given in Eq.~(\ref{eq:coef_dim6}) with $\Lambda$ being replaced
by $\sqrt{\mu_S^{2}}$, and $A^{\rm ND}$ is given by
$A^{\rm ND}  \simeq 0.182{\rm TeV}$.
In Fig.~\ref{fig:ONsinglet_coefficient}, the numerical value of $A(\mu_S^2, \lambda_{\Phi S})$
is shown as a function of $\sqrt{\mu_S^{2}}$ with its limiting values.

The meaning of the value $A^{\rm ND}$ can be better understood if we consider
the case with classically scale invariance (CSI) for the same particle content\cite{Endo:2015ifa,Hashino:2015nxa}.
In this model, all the fields are massless at tree level.
The EWSB occurs at one-loop level by the Coleman-Weinberg mechanism\cite{Coleman:1973jx}.
An interesting prediction is $\Delta \lambda_{hhh}^{\rm CSI}/\lambda_{hhh}^{\rm SM} = +2/3 \simeq +67\%$\cite{Hashino:2015nxa}, which is
independent of $N$.
This universal prediction is the most distinctive feature of the model based on CSI broken by the EWSB.
The sphaleron energy can be calculated by replacing $V(\Phi)$ by the one-loop
effective potential $V_{\rm eff}^{\rm CSI}(\Phi, S_i)$.
We obtain the deviation in the sphaleron energy as $\Delta E_{\sf sph}^{\rm CSI} = -0.121{\rm TeV}$
($\Delta E_{\sf sph}^{\rm CSI}/E_{\sf sph}^{\rm SM} = -1.33$\%).
Taking into account the above prediction on $\Delta \lambda_{hhh}^{\rm CSI}$,
we obtain $A^{\rm CSI} = A^{\rm ND}$.

Finally we discuss the MCHMs, where
the Higgs boson is regarded as the pseudo Nambu-Goldstone boson for the spontaneous breaking
of the global symmetry $SO(5) \to SO(4)$\cite{Kaplan:1983fs}. The sphaleron energy for this model
was first evaluated in Ref.~\cite{Spannowsky:2016ile} for the model with the matter representation to be 4 (MCHM4).
We here discuss the models with two matter representations; i.e., MCHM4 and MCHM5\cite{Carena:2014ria}.
We note that, in these models, the sphaleron energy can be affected not only by
the change in the Higgs potential but also by the deformation of
the term of the covariant derivatives.

For the MCHM4 the Higgs sector is replaced by
\begin{eqnarray}
({\cal D}_i \Phi)^\dagger {\cal D}_i \Phi
  &\to& \frac{1}{4 \Phi^\dagger \Phi}[\partial_i (\Phi^\dagger \Phi)]^2 \left( 1 - \frac{1}{{\cal N}^2}\sin^2 {\cal N} \right)
  \nonumber\\
  &&  + \frac{1}{{\cal N}^2}\sin^2 {\cal N} \;\; ({\cal D}_i \Phi)^\dagger {\cal D}_i \Phi,
  \\
  V(\Phi)  &\to&  \alpha \cos {\cal N}  - \beta \sin^2 {\cal N},
\end{eqnarray}
where $f_\pi$ is the decay constant of the symmetry breaking, $\alpha$ and $\beta$ are constants, and
${\cal N} = \sqrt{2 \Phi^\dagger \Phi}/f_\pi$.
Expanding the energy functional with the inverse of the decay constant $1/f_\pi$ and using the hedgehog ansatz, we obtain
\begin{eqnarray}
&& \hspace{-6mm} \Delta E_{\sf sph}^{\rm MCHM4} = \frac{\pi v^4}{12 f_\pi^2 m_W}\int_0^\infty dr
\left[       r^2 \kappa^2 (3+S^4)\right. \nonumber\\
&& \hspace{-6mm} \times (1-S^2)   \left. - 4S^4 (1-R)^2
  \right] + {\mathcal O}\left(\frac{v^5}{f_\pi^4}\right),
\end{eqnarray}
with $R=R(r)$ and $S=S(r)$ being given in Fig.~\ref{fig:SR}.
We have confirmed that in the case of MCHMs it is valid in $f_\pi>600$GeV that using the sphaleron configuration in the SM as an approximation.
The deviation in the triple Higgs coupling is given by
  $\Delta \lambda_{hhh}^{\rm MCHM4}/\lambda_{hhh}^{\rm SM} \simeq - v^2/(2 f_\pi^2)$.
We then obtain
  $A^{\rm MCHM4}  \simeq 1.939{\rm TeV}$,
which comes from contributions of $\Delta V$ ($+2.918$TeV)
and the covariant derivative term ($-0.979$TeV).
For $f_{\pi}=600$GeV (1TeV), we obtain
$\Delta \lambda_{hhh}^{\rm MCHM4}/\lambda_{hhh}^{\rm SM}=-8.4$\%($-3.3$\%) and
$\Delta E_{\sf sph}^{\rm MCHM4}/E_{\sf sph}^{\rm SM}= +1.78$\% ($+0.647$\%).

For the MCHM5, the covariant derivative term takes the same form as MCHM4,
while the potential differs to
\begin{eqnarray}
  V(\Phi) =  \alpha' \cos ^{2} {\cal N}
                +\beta' \cos ^{2} {\cal N}
                \sin ^{2} {\cal N},
\end{eqnarray}
where $\alpha'$ and $\beta'$ are constants. It gives $\Delta \lambda_{hhh}^{\rm MCHM5}/\lambda_{hhh}^{\rm SM} \simeq - 3v^2/(2 f_\pi^2)$, which is larger
than $\Delta \lambda_{hhh}^{\rm MCHM4}/\lambda_{hhh}^{\rm SM}$ for the same value of $f_\pi$.
Using the same method as the MCHM4, we obtain
  $A^{\rm MCHM5} \simeq 0.745{\rm TeV}$.
For $f_{\pi}=600$GeV  (1TeV), we evaluate
$\Delta \lambda_{hhh}^{\rm MCHM5}/\lambda_{hhh}^{\rm SM}=-25.2$\% ($-9.09$\%) and
$\Delta E_{\sf sph}^{\rm MCHM5}/E_{\sf sph}^{\rm SM}= +2.06$\% ($+0.746$\%).

\section{Discussions and Conclusions}

Several comments are given in order.
First, although the magnitudes of $\Delta E_{\sf sph}^{\rm new}/E_{\sf sph}^{\rm SM}$
discussed in this Letter are at most a few percent, such deviations become crucial
around the transition temperature by the enhancement.
Contrary to the case of the SM\cite{Braibant:1993is}, there are large temperature
dependences in $E_{\sf sph}^{\rm new}(T)$ when the first order phase transition is strong enough\cite{Funakubo:2009eg}.
The sphaleron decoupling condition is sensitive to a small deviation in $\Delta E_{\sf sph}^{\rm new}$ at zero temperature \cite{Chiang:2017nmu,Zhou:2019uzq}.
Therefore, our analysis is significant in connection with the EW baryogenesis.
We explain more details in Appendix A.
Second, the triple Higgs boson coupling is expected to be measured by about 10\% or better accuracies at future high-energy colliders
such as the 0.5-1TeV option of the International Linear Collider\cite{Fujii:2019zll},
the Compact Linear Collider\cite{CLIC:2016zwp},
the High-Energy LHC\cite{Abada:2019ono} and so on.
Third, our analysis can also be applied to the case where multiple VEVs appear in extended Higgs models, as
seen in Ref.~\cite{Fuyuto:2014yia} for the model with the mixing with the singlet scalar field.
Finally, we have focused the relation between $\Delta E_{\sf sph}^{\rm new}$ and $\Delta \lambda_{hhh}^{\rm new}$ since
we are interested in the effect of extended Higgs sectors. For models which affect
the covariant derivative term, the relation with $hWW$ and $hZZ$ couplings would also be interesting.

We have discussed the relation between $\Delta E^{\rm new}_{\sf sph}$ and $\Delta \lambda_{hhh}^{\rm new}$,
which have opposite sign. Our results would be useful to determine the sphaleron energy
by measuring the $hhh$ coupling at future high-energy experiments.

\section*{Acknowledgments}
S. K. was supported, in part, by JSPS, Grant-in-Aid for Scientific Research,
18F18321, 16H06492, 18H04587, 18F18022 and 20H00160.

\appendix
\section{Comments on the sphaleron decoupling condition}
\renewcommand{\theequation}{A.\arabic{equation}}

We here discuss the relation between Eq. \eqref{eq:relation} and the sphaleron decoupling condition which is required for EW baryogenesis.

The sphaleron decoupling condition is given by\cite{Gan:2017mcv,Arnold:1987mh}
\begin{eqnarray}
  \label{eq:sph_decoupling}
  \frac{v(T)}{T}
  &>& \frac{g}{4 \pi \mathcal{E}(T)}
  \left[
    42.8 + 7 \ln \frac{v(T)}{T} - \ln \frac{T}{100{\rm GeV}}
  \right] \nonumber \\
  &\equiv& \zeta_{\sf sph}(T),
\end{eqnarray}
where $\mathcal{E}(T)=E_{\sf sph}(T)/(4 \pi v(T)/g)$.
Cleary, $\zeta_{\sf sph}(T)$ depends on the details of models.
In order to briefly demonstrate the importance of our results on the sphaleorn energy at zero temperature, for an example, we focus on the model with the dimension-six operator in the Higgs potential\cite{Gan:2017mcv,Grojean:2004xa}.
The sphaleron decoupling condition in this model was evaluated by Zhou et al.\cite{Zhou:2019uzq}.
Here, we clarify the relation between the deviation in the sphaleron energy at the zero temperature and the sphaleron decoupling condition.

Requiring the critical temperature $T_C$ and $v(T_C)$ are positive, we obtain the constraint on $\Lambda$\cite{Grojean:2004xa},
\begin{eqnarray}
  \frac{v^2}{m_h} < \Lambda < \frac{\sqrt{3} v^2}{m_h}
  ~
  \Rightarrow
  ~
  484~{\rm GeV} < \Lambda < 839~{\rm GeV}.
\end{eqnarray}
We have caluculated $\zeta_{\sf sph}^{\rm dim6}(T_C)$ at each cutoff scale, and listed the results in Table \ref{table1}.
Also, we have confirmed that the condition given in Eq. \eqref{eq:sph_decoupling} demands $\Lambda<750$GeV which is consistent with the previous analysis\cite{Zhou:2019uzq}.
In the model with $\Lambda=700$GeV, although the deviation in the sphaleron energy at the zero temperature is only $-1.54\%$, while that at around the critical temperature becomes $-19.4\%$.
Therefore, $\zeta_{\sf sph}^{\rm dim6}(T_C)$ is larger than the SM one about $8.5\%$ in the model with the dimension-six operator with $\Lambda=700$GeV.

\begin{table}[t]
  \caption{Sphaleron energy and $\zeta_{\sf sph}(T)$ at the critical temperature.
  $\zeta_{\sf sph}^{\rm SM}(T)$ and $\zeta_{\sf sph}^{\rm dim6}(T)$ indicate the criterion of the sphaleron decoupling condition in the SM and the model with the dimension-six operator, respectively.}
    \label{table1}
  \begin{center}
    \begin{tabular}{|c||c|c|c|c|}
    \hline
     $\Lambda~[{\rm GeV}]$
     & 550
     & 600
     & 650
     & 700
     \\
    \hline \hline
    $\Delta E_{\sf sph}^{\rm dim6}/E_{\sf sph}^{\rm SM}(T=T_C)$
    & $-2.97\%$
    & $-5.78\%$
    & $-11.1\%$
    & $-19.4\%$
    \\
    \hline
    $\mathcal{E}^{\rm dim6}(T=T_C)$
    & 1.806
    & 1.778
    & 1.751
    & 1.723
    \\
    \hline
    $\Delta E_{\sf sph}^{\rm dim6}/E_{\sf sph}^{\rm SM}(T=0)$
    & $-2.65\%$
    & $-2.13\%$
    & $-1.87\%$
    & $-1.54\%$
    \\
    \hline
    $\mathcal{E}^{\rm dim6}(T=0)$
    & 1.865
    & 1.876
    & 1.880
    & 1.886
    \\
    \hline
    $\zeta_{\sf sph}^{\rm dim6}(T=T_C)$
    & 1.50
    & 1.47
    & 1.44
    & 1.41
    \\
    \hline
    $\zeta_{\sf sph}^{\rm SM}(T=T_C)$
    & 1.42
    & 1.36
    & 1.33
    & 1.30
    \\
    \hline
    $v(T_C)/T_C$
    & 3.70
    & 2.80
    & 2.25
    & 1.80
    \\
    \hline
    $\Delta \lambda_{hhh}^{\rm dim6}/\lambda_{hhh}^{\rm SM}$
    & 155\%
    & 131\%
    & 111\%
    & 96\%
    \\
    \hline
    \end{tabular}
  \end{center}
\end{table}
The origin of the enhancement of the deviation in the sphaleron energy is the temperature dependence of $\mathcal{E}(T)$.
In the standard model, $\mathcal{E}^{\rm SM}(T)$ is a constant ($\mathcal{E}^{\rm SM}(T)=1.92$)\cite{Braibant:1993is}.
On the other hand, as shown in the literature \cite{Chiang:2017nmu,Zhou:2019uzq}, $\mathcal{E}(T)$ decreases with an increase in $T$ in many of extended Higgs models.
Then, the temperature dependence of $\zeta_{\sf sph}(T)$ comes not only from $v(T)$ but also from $\mathcal{E}(T)$.
In some cases, a few percent deviation in the sphaleron energy at the zero temperature corresponds to 10\% deviation at the critical temperature such as the model with $\Lambda=700$GeV.
Consequently,  $\zeta_{\sf sph}(T)$ at the critical temperature is larger than the SM value.

As discussed above, the deviation in the sphaleron energy at the finite temperature can be large even if the deviation at the zero temperature is a few percent.
Therefore, in order to verify the scenario of the EW baryogenesis, we need to determine the sphaleron energy at the zero temperature precisely.
This can be achieved by detecting the deviation in the triple Higgs boson coupling at future colliders.


\end{document}